\documentclass{ws-ijmpd}
\usepackage{amsmath,amssymb,slashed}
\usepackage[hyperfootnotes=false]{hyperref}
\newcommand{\be}{\begin{equation}}
\newcommand{\ee}{\end{equation}}
\newcommand{\bea}{\begin{eqnarray}}
\newcommand{\eea}{\end{eqnarray}}

\newcommand{\CMP}{{\it Commun. Math. Phys.\,}}

\newcommand{\JMP}{{\it J. Math. Phys.\,}}

\newcommand{\NP}{{\it Nucl. Phys.\,}}

\newcommand{\PR}{{\it Phys. Rev.\,}}

\begin{document}

\markboth{Chen, Chiang and Hu }
{Instructions for Typing Manuscripts (Paper's Title)}

%
\catchline{}{}{}{}{}
%

\title{A Quantized Spacetime Based on $Spin(3,1)$ Symmetry}

\author{Pisin Chen$^{a1234}$,Hsu-Wen~Chiang$^{b13}$ and Yao-Chieh Hu$^{c123}$}

\address{$^{1}$Department~of~Physics, National~Taiwan~University, Taipei~10617, Taiwan, R.O.C.\\
	$^{2}$Graduate~Institute~of~Astrophysics, National~Taiwan~University, Taipei~10617, Taiwan, R.O.C.\\
	$^{3}$Leung Center for Cosmology and Particle Astrophysics, National~Taiwan~University, Taipei~10617, Taiwan, R.O.C.\\
	$^{4}$Kavli~Institute~for~Particle~Astrophysics~and~Cosmology, SLAC~National~Accelerator~Laboratory, Stanford~University, Stanford, CA~94305, U.S.A.\\
	$^{a}$E-mail: pisinchen@phys.ntu.edu.tw\\
	$^{b}$E-mail: b98202036@ntu.edu.tw\\
	$^{c}$E-mail: r04244003@ntu.edu.tw\\
}

\maketitle


\begin{abstract}
We introduce a new type of the spacetime quantization based on the spinorial description suggested by loop quantum gravity. Specifically, we build our theory on a string theory inspired $Spin(3,1)$ worldsheet action. Because of its connection with quantum gravity theories, our proposal may in principle link back to string theory, connect to loop quantum gravity where $SU(2)$ is suggested as the fundamental symmetry, or serve as a Lorentzian spin network. We derive the generalized uncertainty principle and demonstrate the holographic nature of our theory. Due to the quantization of spacetime, geodesics in our theory are fuzzy, but the fuzziness is shown to be much below conceivable astrophysical bounds.
\end{abstract}
\keywords{Non-commutative geometry; quantum gravity; generalized uncertainty principle.}
\ccode{PACS numbers: 04.60.-m. }

\section{Introduction}
\label{sec:intro}
In both superstring and loop quantum gravity theories, there exist a minimal distance measure, which suggests that the infinitely-differentiable spacetime manifold may be an illusion that ceases to be valid at very small scales. This suggests that spacetime may be quantized. Snyder coined the attempt as ``quantized spacetime".\cite{Snyder} Motivated by possible fundamental descriptions of the quantum gravity, we introduce a new type of the spacetime quantization based on the spinorial description, specifically, a string theory inspired $Spin(3,1)$ worldsheet action.\cite{CHC}

In the attempts of quantizing spacetime, the quadratic distance function $g$ is usually promoted to its quantum version without modification. A new way of deforming the spacetime algebra was recently proposed by Adler.\cite{Adler1} Instead of taking the spectral square root\cite{LQGlength0} of $g$ to obtain the spectrum of the distance operator, this new approach takes the route via the Clifford algebra. One then obtains a linear line element $\Delta s_L = \gamma _\mu \Delta x^\mu$. The price to pay, however, is that the new transfer function $\gamma _\mu$ is a matrix instead of a pure number. To compensate that, Adler reinterprets $\Delta s_L$ as a quantum mechanical (QM) object $\hat{\Delta s}$ called ``linear line element operator", whose eigenvalue is the proper distance and the state describes an eigendirection $\epsilon _\mu$ along which the measure is completely certain, i.e.,
\begin{align}
\Delta s = \langle \hat{\Delta s} \rangle = \left\langle \epsilon \left| \gamma _\mu \Delta x^\mu \right| \epsilon \right\rangle = \left( \left| \epsilon \right\rangle \right) ^\dagger \gamma _\mu \Delta x^\mu \left| \epsilon \right\rangle = \epsilon _\mu \Delta x^\mu \,. \label{ds_A}
\end{align}
The proper distance is thus treated as the addition of generators of the spin group on the classical $x$-space, similar to the 't Hooft operator on a spacetime lattice, and the infinitely differentiable manifold is replaced by a piecewise-linear one.
\footnote[0]{For the sake of readability, $IJK$ are for $SO(3,1)$ tangent bundle coordinate indices, $\mu \nu \lambda \rho$ for 4-$d$ coordinate, $i j k$ for 3-$d$ coordinate, and $(i)(j)(k)$ for site numbers. Gamma matrices $\gamma$ are defined according to the type of the indices (Minkowskian if unspecified). We also use natural units $c = \hslash = G = 1$, and the signature of $g$ is chosen to be $\left(+,-,-,-\right)$.}

An uncertainty is introduced as $var(\Delta s) = \Delta s^2 - { \left( \dot{x}^\mu \epsilon _\mu \right) }^2 \Delta s^2$, where $\dot{x}^\mu$ is the four-velocity. The preferred direction $\epsilon _\mu$ along which the measurement is completely certain inevitably breaks the Lorentz invariance, unless the theory is nondissipative or is at high temperature limit. The whole purpose of using the Clifford algebra to preserve the Lorentz invariance is therefore defeated. Another problem is that gamma matrices and the coordinate difference are living in different representations, rendering the linear line element representation dependent. In next section, we will show that following Adler's philosophy but recasting the linear line element as $\Delta s = \gamma _\mu \left\langle \lambda \gamma ^\mu \right\rangle$, a new kind of spacetime quantization can be developed that is fully covariant, representation independent, and with rich structures and salient properties.\cite{CHC}

\section{Constructing Action through Bosonization}
\label{sec:action}

Equation \eqref{ds_A} suggests that Adler's approach of spacetime quantization may be recast as a worldline action in terms of bi-fermionic fields that exhibit bosonic behavior. A natural tool to attain this action would be the bosonization in 2$d$ quantum field theory (QFT).\cite{bosonization} By considering the worldline as the low energy limit of a cylindrical worldsheet, we may apply the fermionization on the worldsheet action and compactify it back to 1$d$. To fix the gauge structure $g(X)$ such that the fermionization procedure can be applied, we utilize the tetrad formalism and obtain a $SO \left( 3,1 \right)$ charged scalar theory. However, since we are aspired to construct a spinor theory, we replace the group structure with $Spin(3,1)$, which is effectively equivalent to $SO(3,1)$ classically. 

We fermionize the theory by checking the bosonization dictionary\cite{nonabelianbosonization} and arrive at a trivial charged fermion action describing a $4$d spacetime spinor valued $2$d worldsheet Majorana-Weyl spinor. From constraints of the worldline action, we can finally obtain the compactified action  in terms of the spinor field $\psi$. Since the worldsheet spinor is conserved, we obtain the quantization condition\cite{CHC} $\Delta X^I = {\lambda v^I}/{\sqrt{\left| v^2 \right|}}$, where $v^I \equiv d X^I / d \tau$ is the velocity of the original worldline action in terms of tetrads and $\lambda \sim \mathcal{O}(\text{Planck length})$ is the characteristic length of the quantized spacetime. We can also show that $\big{\langle} \Delta \hat{X}^\mu \big{\rangle} = \lambda \bar{\psi} \gamma ^\mu \psi = \lambda v^\mu$, where $v^2 = 1, 0, {\rm or} -1$. Notice that since here we focus on eigenstates of the velocity operator, $ds^2 = \bar{\psi} \psi$ must vanish. If the constraint is substituted with $ds^2 = {\rm const.}$, we immediately obtain $\Delta s^2 = \bar{\psi} \psi = 1, 0, {\rm or} -1$. We now have a concrete quantization scheme that can be reproduced explicitly by a fermionized worldline action.

\section{Spacetime Interval Operator}
Let us introduce the ``spacetime interval operator" $\Delta \hat{X}^I$ as
\begin{align}
\big{\langle} \Delta \hat{X}^I \big{\rangle} = \bar{\psi} \Delta \hat{X}^I \psi = \bar{\psi} e^I_\mu \Delta \hat{X}^\mu \psi &= \psi ^{\dagger} \gamma ^0 \Delta \hat{X}^I \psi = \lambda \psi ^{\dagger} \gamma ^0 \gamma ^I \psi \,, \label{measure} \\
\Delta s^2 = \eta _{IJ} \left\langle \Delta \hat{X}^I \Delta \hat{X}^J \right\rangle &= 4\lambda ^2 \bar{\psi} \psi = \pm 4\lambda ^2, {\rm or}\ 0 \,. \label{normalization}
\end{align}
Here $e^I_\mu$ is the tetrad. $\big{\langle} \Delta \hat{X}^I \big{\rangle}$ is a normalized non-null vector for a non-null state or a null vector with positive time component for a null state. The choice of the normalization can be appreciated by looking at the solution of the Dirac field equation.

This new measurement is neither a distance measure nor a coordinate difference operator, but a difference operator on the tangent bundle of the manifold. A better way to understand it is to treat it as a discretized version of the velocity four-vector. Only when combined with the tetrad does the coordinate difference measure $\Delta \hat{X}^\mu$ reappear. Just like a vector in GR, the components of $\Delta \hat{X}^I$ are not physical. To measure the velocity of a particle we need a two-particle interaction, and the physical object is the inner product $r_I \Delta \hat{X}^I$, where $r_I$ is the classical trajectory of a probe particle expressed in the same representation as $\Delta \hat{X}^I$. Therefore all the derivations and results we obtained are representation independent. 

The new choice of interpretation also implies the existence of an underlying minimal distance. But unlike what Adler obtained\cite{Adler1}, in our case even uncertainties are Lorentz invariant. One may try to obtain the variances of the interval measures
\bea
\left\langle \text{var} \left( r_I \Delta X^I \right) \right\rangle &=& \left| \left\langle {\left( r_I \Delta X^I \right) }^2 \right\rangle -{\left\langle r_I \Delta X^I \right\rangle }^2 \right| = \lambda ^2 \left| r_I r^I n_J n^J - {\left( r_K n^K \right) }^2 \right| \,, \label{var_dX} \\
\left\langle \text{var} \left( \Delta s \right) \right\rangle &=& \left| \left\langle \Delta X^I \Delta X_I \right\rangle - \left\langle \Delta X^I \right\rangle \left\langle \Delta X_I \right\rangle \right| = 3 \lambda ^2 \left| n_I n^I \right| \,. \label{var_ds}
\eea
Here $n^I = \left\langle \Delta X^I \right\rangle$ and $r^I$ is the ruler. Clearly, along the eigen-direction there is no uncertainty. However, along the transverse direction the measurement is completely uncertain. From this point of view, the behavior of the spacetime interval operator is exactly the same as the spin operators in relativistic QM. 

One interesting feature of our theory is that from equation \eqref{measure}, the time difference is proportional to the identity operator, rendering its expectation value positive definite. Thus the arrow of time problem is a nonissue without the need to invoke the second law of thermodynamics.

\section{Holography, Non-commutativity and Generalized Uncetainty Principle}
\label{sec:commutator}
We can define the position operator and the position eigenstate as $\left\langle \hat{X}^\mu \right\rangle = \sum \lambda \bar{\psi }_{ \left( i \right) } \gamma ^\mu \psi _{ \left( i \right) }$. The position operator can be envisioned as the 't Hooft operator on a 1$d$ spin chain, and the associated Fourier transformation operator as the Wilson line operator. This kind of visualization is inherited from the bosonization. It is now straightforward to derive the commutation and anti-commutation relations:
\begin{align}
\bigg{\langle} \big{[}  \hat{X}^\mu &, \hat{X}^\nu \big{]} \bigg{\rangle} = \sum _i \bigg{\langle} \big{[} \Delta \hat{X}^\mu _{\left( i \right)} , \Delta \hat{X}^\nu _{\left( i \right)} \big{]} \bigg{\rangle} \,,\\
\frac{1}{2} \Bigg{\langle} \bigg{\{ } \hat{X}^\mu &, \hat{X}^\nu \bigg{\} } \Bigg{\rangle} = \left\langle \hat{X}^\mu \right\rangle \left\langle \hat{X}^\nu \right\rangle + \sum _i \left( \lambda ^2 g^{\mu\nu} - \left\langle \Delta X^\mu \right\rangle _{\left( i \right)} \left\langle \Delta X^\nu \right\rangle _{\left( i \right)} \right) \,.
\end{align}
Clearly the spacetime in this theory, contrary to those of most others, is neither commutative nor anti-commutative.

One important aspect of our theory is its holographic nature. Since a $N$-line-element state resembles a composite spin-$N/2$ charged particle, by Clebsch-Gordan decomposition it is with no doubt that only $\mathcal{O}\left(N^3\right)$ different states are allowed within a four-hyperbola, implying that the spacetime information can be written in a three-dimensional quantum language. With the geodesic equation\cite{CHC}, the degree of freedom, i.e., the entropy, is proportional to the surface area of the system.

Another aspect is that it implies the generalized uncertainty principle (GUP). Let us invoke the ADM formalism to reduce the symmetry down to $Spin(3)$. Under this symmetry, the spacetime behaves as a spin chain in the zero temperature limit, with the coordinate measure being the total spin. It can be reinterpreted as a bosonic system with $SU(2)$ symmetry even at the quantum level, as demonstrated in the non-abelian bosonization. An immediate conclusion is the existence of an exterior derivative that can be treated as a momentum operator.\cite{CHC,SU2F}
One can therefore compute the uncertainty relation, and in the low momentum limit and $\left\langle x \right\rangle \to 0$ we obtain the GUP in the usual form:
\bea
\Delta x_i \Delta p_j  \geqslant \frac{1}{2} \left( \delta _{ij} + 2 \lambda ^2 \delta _{ij} \left\langle p_i \right\rangle \left\langle p_j \right\rangle \right) \,.
\eea

\section{Astrophysical Tests}
\label{sec:astrophystest}
It is straightforward to apply the quantized geodesic obtained in the previous section as an astrophysical test for this theory through the variance of the geodesic, i.e., the smearing effect. 
Consider, for example, the null geodesics of photons with characteristic energy $E\sim 100$ keV emitted from a gamma ray burst (GRB) afterglow at high redshift ($z \approx 10$). Because the Lorentz invariance is manifest in our theory, the smearing effect occurs only along the perpendicular direction. Assuming a Gaussian distribution of photons at the source, we find, from equations \eqref{var_dX} and \eqref{var_ds}, that the fuzziness of the GRB image so induced upon arrival on earth is of the order of $\sqrt{ \lambda ^2 E d_S \left( z \approx 10 \right) } \approx {10}^{-13}$ meter, where $d_S = \int _0^z dz'(1+z')/H(z')$ is the rescaled distance taking the rescaling of the variance into account. Clearly, it is impossible to identity such minute effect in the near future.
\section{Conclusion}
We show that a non-commutative spacetime theory can be constructed on top of Adler's ``linear line element", and derive the associated action from the fermionization of the K-K ground state of the Polyakov action on a cylinder. The theory is shown to be holographic, and passes the astrophysical tests on the smearing effect and Lorentz invariance violation. The theory can also be regarded as the $4$-D extension to the $SU(2)$-type models, and from there the generalized uncertainty principle is derived.

This theory is quite potent, with many additional implications that deserve further exploration. For example, the coefficient $\xi$ of the Bekenstein-Hawking entropy, $S = \xi A / 4$, may be fixed in our setup. Also so far our theory is still purely kinematic. The dynamics of the theory can be introduced by either invoking loop quantum gravity or by putting back the K-K tower of the worldsheet excitations. Interestingly in loop quantum gravity\cite{LQGlength2}, it has been argued that the length operator should be defined through Dirac operators. It remains curious whether our work can be served as a representation that can describe length, area and volume easily without introducing the spectral decomposition.\cite{LQGlength0,LQGlength1}
	\section*{Acknowledgement}
	We appreciate useful discussions with Abhay Ashtekar, Ron Adler and Carlo Rovelli. This work is supported by National Center for Theoretical Sciences (NCTS) of Taiwan, Ministry of Science and Technology (MOST) of Taiwan, and the Leung Center for Cosmology and Particle Astrophysics (LeCosPA) of National Taiwan University.


\begin{thebibliography}{99}
	
	\bibitem{Snyder}
	H. S. Snyder,
	``Quantized Space-Time'',
	\href{http://journals.aps.org/pr/abstract/10.1103/PhysRev.71.38}{{\PR} {\bf 71}, 38 (1947)}.
	
	\bibitem{CHC}
	H.-W. Chiang, Y.-C. Hu, P. Chen,
	``Quantization of Spacetime Based on Spacetime Interval Operator",
	Phys. Rev. D 93, 084043 (2016). 
	\href{http://arxiv.org/abs/1512.03157}{[arXiv:1512.03157]}.
	
	\bibitem{Adler1}
	R. J. Adler,
	``A quantum theory of distance along a curve'',
	unpublished (2014)
	\href{http://arxiv.org/abs/1402.5921}{[arXiv:1402.5921]}.
	
	\bibitem{LQGlength0}
	T. Thiemann,
	``A length operator for canonical quantum gravity'',
	{\JMP} {\bf 39}, 3372-3392 (1998),
	\href{http://arxiv.org/abs/0806.4710}{[arXiv:gr-qc/9606092]}.
	
	\bibitem{bosonization}
	S. Coleman,
	``Quantum sine-Gordon equation as the massive Thirring model'',
	{\PR} {\bf D11}, 2088 (1975),
	\href{http://users.physik.fu-berlin.de/~kamecke/ps/coleman.pdf}{Freie Universit\"at Berlin}.
	
	\bibitem{nonabelianbosonization}
	E. Witten,
	``Non-Άbelian Bosonization in Two Dimensions'',
	{\CMP} {\bf 92}, 455-472 (1984),
	\href{https://projecteuclid.org/euclid.cmp/1103940923}{Project Euclid}.
	
	\bibitem{SU2F}
	E. Batista and S. Majid,
	``Noncommutative Geometry Of Angular Momentum Space $U(\mathfrak{s}u(2))$'',
	{\JMP} {\bf 44}, 107 (2003)
	\href{http://arxiv.org/abs/hep-th/0205128}{[arXiv:hep-th/0205128]}.
	
	\bibitem{LQGlength2}
	C. Rovelli,
	``Lorentzian Connes Distance, Spectral Graph Distance and Loop Gravity'',
	Unpublished (2014)
	\href{http://arxiv.org/abs/1408.3260}{[arXiv:1408.3260]}.
	
	\bibitem{LQGlength1}
	E. Bianchi,
	``The length operator in Loop Quantum Gravity'',
	{\NP} {\bf B807}, 591 (2009)
	\href{http://arxiv.org/abs/0806.4710}{[arXiv:0806.4710]}.
	
\end{thebibliography}
\end{document}